# Phosphates reveal high pH ocean water on Enceladus

Christopher R. Glein [*] and Ngoc Truong

Space Science Division, Space Sector, Southwest Research Institute, 6220 Culebra Road, San Antonio, TX 78238-5166, United States

[*] Corresponding author. Email: christopher.glein@swri.org

## Highlights

- pH is one of the most important parameters characterizing natural waters.
- The observed phosphate speciation constrains the pH of Enceladus's ocean.
- A high pH of 10.1-11.6 is inferred, with a most consistent value of ~10.6.
- Equilibrium exsolution can link the plume's $CO_2/H_2O$ ratio to the pH from phosphates.
- Water-rock interactions most likely led to a high pH ocean on Enceladus.

## Abstract

Enceladus offers our best opportunity for exploring the chemistry of an ocean on another world. Here, we perform geochemical modeling to show how the distribution of phosphate species found in ice grains from Enceladus's plume provides a very straightforward constraint on the pH of the host solution. The ratio of $HPO_4/PO_4$ species serves as a pH indicator. We find evidence of moderately alkaline water (pH 10.1-11.6)—significantly more alkaline than current estimates (~8-9) of the pH of Enceladus's ocean. Nevertheless, the pH range from phosphates is consistent with the $CO_2/H_2O$ ratio measured in the plume if $CO_2$ exsolves from ocean water according to its equilibrium solubility. A simple energy balance can be used to quantify volatile fractionation during gas transport inside Enceladus's tiger stripes; we deduce that ~83% of water vapor is removed as ice during transport between the liquid-vapor interface and where gases exit the subsurface. We also explore how $CO_2$ degassing may lead to an increase in the apparent pH of ocean water. We generate maps of allowed combinations of pH and dissolved inorganic carbon concentration of the source water for a wide range of scenarios. Our preferred interpretation, constrained by the observed heat flux, implies minimal $CO_2$ degassing from ocean water. Hence, the pH recorded by phosphates should closely approximate that of the ocean; our best estimate is pH ~10.6. Such a high pH seems to reflect a major role of silicates enriched in Na, Mg, or Fe(II) interacting extensively with ocean water. Silica nanoparticles would not form or would subsequently dissolve if the pH is too high (>10.5). The outgassing model presented here provides a new path to quantify the dissolved concentrations of volatile species.

**Keywords:** Cosmochemistry; Enceladus; Interiors; Satellites, composition; Volcanism

# 1. Introduction

## 1.1. One variable to rule them all in aqueous geochemistry

The pH is a geochemical parameter that determines the stable forms of dissolved chemical species and the solubilities of minerals. It can also strongly influence the rates of mineral dissolution and the energetics of microbial metabolisms (Jin & Kirk, 2018). In field geochemistry, pH is one of the first measurements that is made to characterize the nature of aqueous solutions. The pH quantifies the acidity of a solution; it is defined as

$$pH = -\log a_{H^+},\tag{1}$$

where $a_{H^+}$ designates the activity of $H^+$ referenced to the ideal dilute standard state on the molality (mol/kg $H_2O$) scale at the temperature and pressure of interest.

## 1.2. pH is integral to Enceladus's ocean chemistry

Saturn's moon Enceladus has an ice-covered ocean that erupts into space, forming a plume of gases and ice grains (Hansen et al., 2006; Porco et al., 2006; Postberg et al., 2009; Thomas et al., 2016; Villanueva et al., 2023). This ocean is thought to be one of our best bets in the search for extraterrestrial life because its composition appears to make it habitable (Waite et al., 2009, 2017; Postberg et al., 2018, 2023; Ray et al., 2021; Xu et al., 2025), and the plume provides samples of the ocean without the need to dig or drill (Cable et al., 2021). Accordingly, considerable effort has gone into understanding the chemistry of Enceladus's ocean (Zolotov, 2007; Glein et al., 2018; Fifer et al., 2022; Ramírez-Cabañas et al., 2024). However, unlike on Earth where we can use a pH meter, we cannot directly determine pH on Enceladus. Various indirect approaches have been developed to estimate the pH, but they yield values spanning from weakly (pH ~8) to strongly (pH ~12) alkaline (Postberg et al., 2009; Glein et al., 2015; Hsu et al., 2015; Glein & Waite, 2020; Fifer et al., 2022).

Recently, a consensus has been building around an ocean pH of ~8-9 (Glein & Waite, 2020; Fifer et al., 2022). These researchers developed geochemical models of ocean chemistry based on different assumptions about how the process of forming the plume from ocean water containing salts and dissolved gases would fractionate materials between the ocean and plume. They then inferred pH ranges that would allow a consistent interpretation of Cassini data using their models. A weakly basic ocean seemed attractive because it was consistent with the pattern of pH-dependent peaks observed in mass spectra of salt-rich plume particles (Postberg et al., 2009), and with a hydrothermal interpretation for signatures of silica seen streaming from the Saturn system (Hsu et al., 2015). However, the analysis of mass spectra of salt-rich plume particles has advanced greatly (Postberg et al., 2021; Seaton et al., 2025) since the pioneering work of Postberg et al. (2009). We will return to the topic of hydrothermal silica in Section 4.2. The prevailing estimates for the pH of Enceladus's ocean may not be as well constrained as commonly perceived. If so, then there is a need for additional data.

## 1.3. Phosphate speciation as a novel indicator of pH

After the studies of Glein & Waite (2020) and Fifer et al. (2022), phosphate salts were discovered in ice grains from the plume. Postberg et al. (2023) found that observed mass spectra of phosphate-rich grains can be reproduced from laboratory mixtures with molar ratios of $Na_2HPO_4/Na_3PO_4$ between 2.5 and 25. They used a laser to disperse ions from a liquid-water beam containing these salts. This approach is meant to mimic impact ionization of salt-bearing ice grains analyzed in space (Klenner et al., 2019). The reported range of $Na_2HPO_4/Na_3PO_4$ ratios gave best matches to the peak pattern of three phosphate-bearing Na-cluster cations identified in mass spectra of nine P-rich grains that were discovered in Saturn's E ring. It is fair to question the relevance of liquid experiments to solid ice particles; this topic is an active area of research (e.g., Burke et al., 2023). Nevertheless, we will interpret the observational constraint on the ratio of phosphate species using models of aqueous chemistry that do not involve brines or the precipitation of salts. A direct comparison between Postberg et al.'s (2023) liquid-based constraint and our models may be most appropriate. In any event, their constraint provides a new opportunity to test our understanding of the pH of Enceladus's ocean.

The phosphate system is a classic example of polyprotic acid-base equilibria. Aqueous phosphoric acid ($H_3PO_4$) is the starting point, followed by a series of acid dissociation reactions that terminate at the phosphate anion ($PO_4^{-3}$):

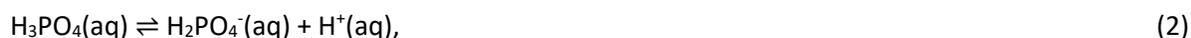
$$H_3PO_4(aq) \rightleftharpoons H_2PO_4^-(aq) + H^+(aq), \tag{2}$$

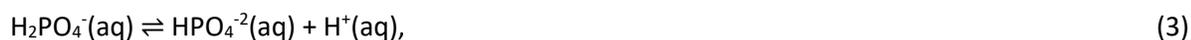
$$H_2PO_4^-(aq) \rightleftharpoons HPO_4^{-2}(aq) + H^+(aq), \tag{3}$$

and

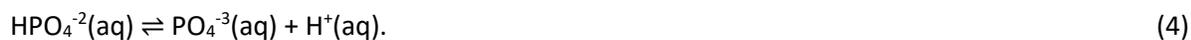
$$HPO_4^{-2}(aq) \rightleftharpoons PO_4^{-3}(aq) + H^+(aq). \tag{4}$$

We can gain first-order insight into how the relative abundances of these phosphate species depend on pH by constructing a Bjerrum plot. Figure 1 shows the behavior for an ideal solution based on thermodynamic data from Shock et al. (1989, 1997). In an ideal solution, all species have activities that equal their molalities (see Section 2). This assumption is known to diverge from reality as solutions become saltier. Nevertheless, starting with an ideal solution can be helpful to illustrate fundamental concepts of aqueous geochemistry. If we proceed on a provisional basis, then it is evident that, at pH ~8-9, the phosphate speciation should be dominated by $HPO_4^{-2}$ and $H_2PO_4^-$, whereas $PO_4^{-3}$ should have a negligible abundance. This expectation is inconsistent with the phosphate speciation determined by Postberg et al. (2023). It appears that the pH needs to be higher to obtain the correct phosphate speciation. However, more detailed quantification using Figure 1 would be ill-advised because these simple calculations do not account for salt effects. Therefore, the main objective of this work is to constrain the pH using a more robust geochemical model of phosphate speciation on Enceladus. We will then explore how representative the "plume pH" is of the ocean below, and conclude with a few of the most important implications of a new pH range.

**Figure 1.** How the ideal distribution of phosphate species changes with pH at 0 °C and 1 bar. The pH of neutral water at these conditions is indicated. Also indicated is the currently accepted range for the pH of Enceladus's ocean (Glein & Waite, 2020; Fifer et al., 2022).

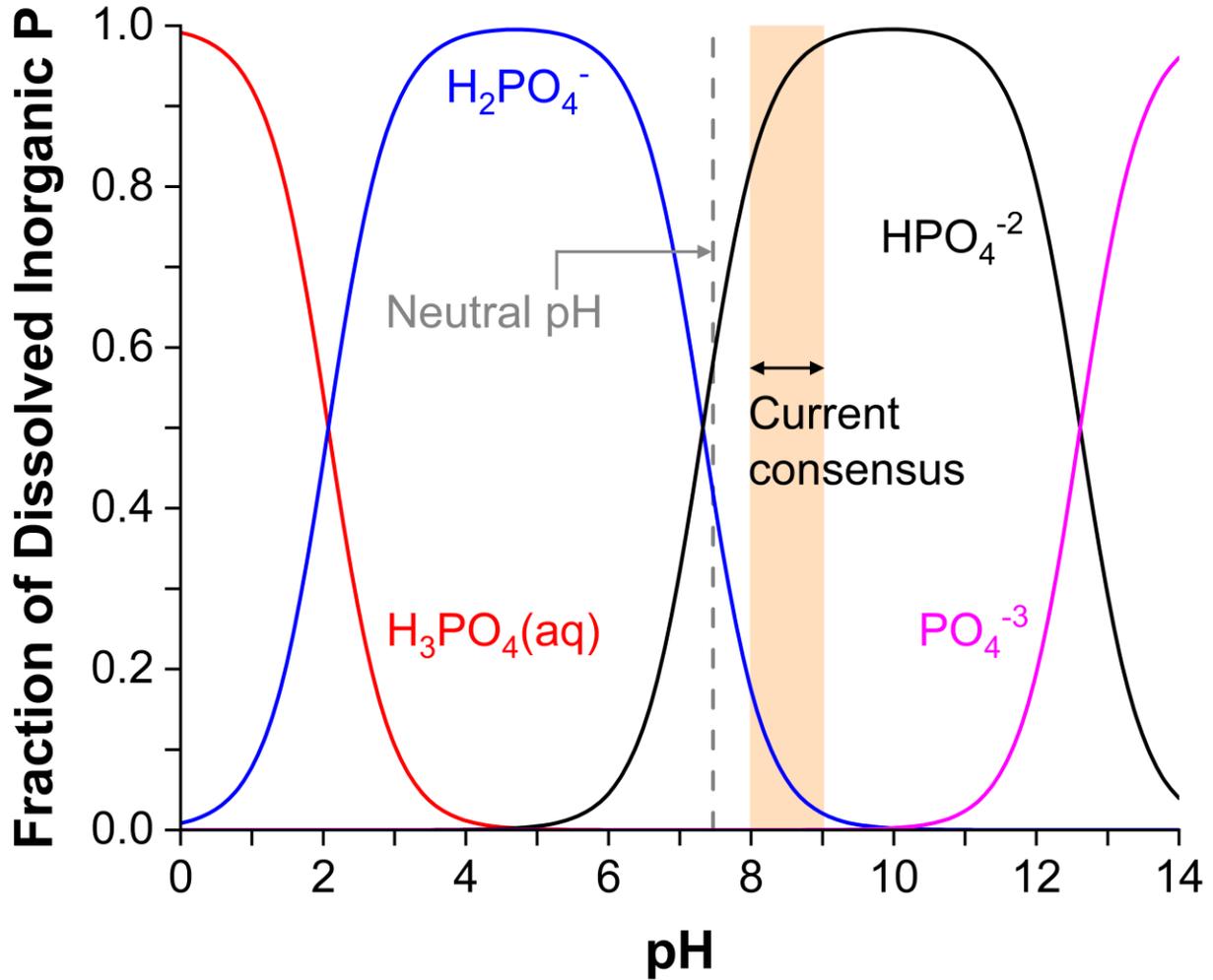

## 2. Modeling approach

Reaction (4) is the key reaction that enables us to relate Postberg et al.'s (2023) constraint on the $Na_2HPO_4/Na_3PO_4$ ratio to solution pH. The law of mass action for the equilibrium constant ($K$) of this reaction can be written as shown below:

$$K_4 = \frac{a_{PO_4^{-3}} a_{H^+}}{a_{HPO_4^{-2}}} = \frac{\gamma_{PO_4^{-3}}}{\gamma_{HPO_4^{-2}}} \frac{m_{PO_4^{-3}}}{m_{HPO_4^{-2}}} a_{H^+}, \qquad (5)$$

where $a_i$, $\gamma_i$, and $m_i$ refer to the activity, activity coefficient, and molality of the $i$th species, respectively. To calculate the pH via Equation (1), we can see that the concentration ratio of phosphate species is necessary but insufficient to evaluate Equation (5)—we must also know the ratio of activity coefficients. The latter is one place where salt effects come into play.

Equation (5) shows that activity coefficients are needed to solve the pH problem, and these correction factors are expected to vary with the salinity of phosphate-bearing fluids. Therefore, to constrain pH, we need to set limits on salinity. Postberg et al. (2009) provided the first constraints on the salt chemistry of plume ice grains, but they have been making improvements since then (Postberg et al., 2021). Based on the latest, publicly presented results, we have defined salt-poor and salt-rich endmembers (Table S1).

Our geochemical model is of the $Na_2O-K_2O-CO_2-H_3PO_4-HCl-HBr-H_2O$ system, which includes all observed salt species (Postberg et al., 2009, 2021, 2023). While the concentration of dissolved Ca can be expected to be an important controller of the concentration of dissolved P, the concentration of dissolved Ca does not need to be specified in a model of aqueous speciation only (i.e., no seafloor minerals) when we already know the concentration of dissolved P (Postberg et al., 2023). To the best of our knowledge, Ca has never been detected in plume or E-ring ice grains from Enceladus—this lack of detection is broadly consistent with the presence of abundant P. We compute the system's speciation over a reaction path of varying pH using the React app in The Geochemist's Workbench 2023. Because salty plume grains will record conditions near the ice-liquid interface, we assume that the freezing temperature of water is most appropriate and the corresponding pressure is low. Our model conditions are 0 °C and 1 bar ($10^5$ Pa) [1]. The overall approach is similar to that of Glein et al. (2015), except here we use an updated thermodynamic database. Our new database was generated from the default database in the PyGeochemCalc package (Awolayo & Tutolo, 2022). We added 13 species to this database that may be important to the speciation on Enceladus: $NaCO_3^-$, $NaHCO_3(aq)$, and $H_2CO_3(aq)$ from the Deep Earth Water model (Huang & Sverjensky, 2019), and a set of Na and K phosphate complexes from Daniele et al. (1991) (see Thermodynamic data file). In our React calculations, activity coefficients of ionic species are computed using an extended Debye-Hückel equation (Helgeson, 1969), since the plume ice is salty but is not a brine.

We perform additional calculations using React to explore the chemical consequences of $CO_2$ degassing from carbonate-bearing fluids on Enceladus. The same thermodynamic database is used. This scenario assumes that pre-degassed ocean water could have a lower pH and higher dissolved inorganic carbon (DIC) concentration than those represented by plume salts. We seek to constrain initial ocean conditions of pH-DIC that would be allowed as $CO_2$ is progressively removed from the source fluid. To keep the focus on these key unknowns, in this set of calculations, we consider intermediate values for the following input parameters: Cl = 0.2 molal, dissolved inorganic phosphorus (DIP) = 0.004 molal (see Table S1), and K/Na = 1/16 (i.e., CI chondritic; Lodders, 2021).

### 3. Results and discussion

*3.1. High pH yields the observed phosphate speciation*

Enceladus's observed phosphate speciation is indicative of a pH between 10.1 and 11.6 (Figure 2). The overall uncertainty due to observed salts is ~0.5 pH units, with the salt-rich endmember shifted to lower pH. It should be noted that our calculations account for the effects of bulk salts measured in frozen ocean water (Postberg et al., 2009, 2011, 2021, 2023). These ice grains are salty but do not appear to be brines. However, one could imagine residual liquid to keep respeciating until freezing is complete. This type of evolution may allow the pH to be lower than what we calculated (Figure 2). On the other hand, salt effects may not be extrapolated so simply into the highly concentrated regime owing to ion association and other complications. There is also the question of how much of the chemical system is kinetically responsive once it is almost totally frozen (Fox-Powell & Cousins, 2021). Such complexity introduces uncertainty that is at present difficult to quantify, but we would be surprised if the pH were to go below ~9. A large change in $a_{H^+}$ would be needed since pH is expressed on a logarithmic scale. Future work is required to test this suspicion. Alternatively, because Postberg et al. (2023) did not use brines when they constrained the $Na_2HPO_4/Na_3PO_4$ ratio, a model of brine chemistry may not be needed to interpret their constraint.

---

[1] The most widely used pressure in thermodynamic databases at temperatures below 100 °C is 1 bar. Nevertheless, equilibria between aqueous species are negligibly affected by a small pressure difference between the triple-point pressure (~6.1 mbar) and 1 bar.

**Figure 2.** Using the observed ratio of $HPO_4$ and $PO_4$ species (Postberg et al., 2023) to deduce the pH of their host solution. Summation signs indicate all dissolved species containing $HPO_4$ or $PO_4$, including ion pairs such as $NaHPO_4^-$. Red and blue lines show ratios that are predicted after the compositions given in Table S1 are speciated as a function of pH at 0 °C and 1 bar. The ideal speciation shown in Figure 1 is represented by the gray dashed line. An ideal solution ignores the effects of salts on ion activities, as ions in such a solution are assumed to behave as if they were in an infinitely dilute solvent. The dark green region shows where both constraints ($\Sigma HPO_4/\Sigma PO_4$ ratio and salinity) converge.

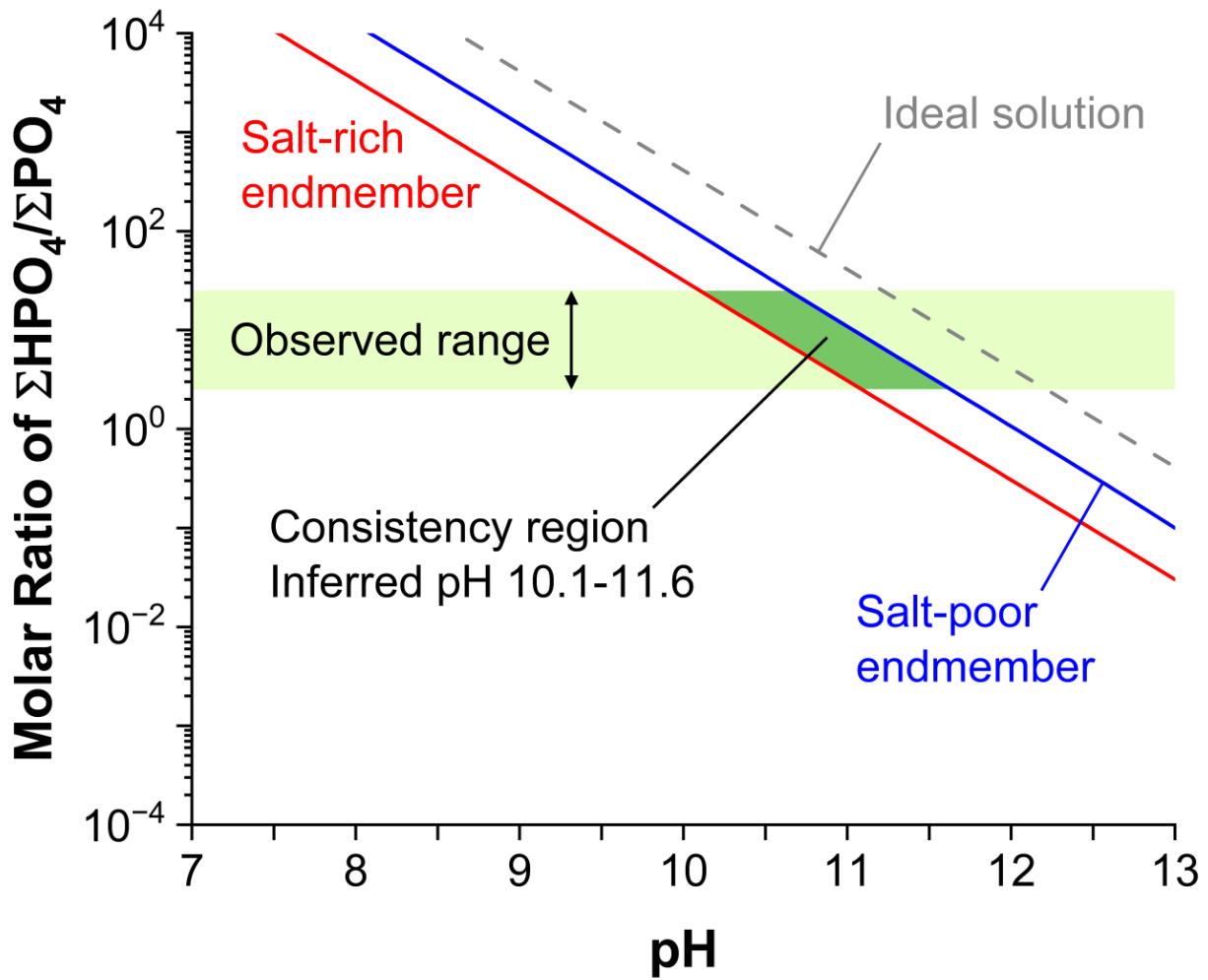

Our speciation calculations provide other details about the chemical nature of Enceladean ocean water. Sodium and chloride ions are found to dominate the distribution of species, and the dominant form of DIC is carbonate in one form or another (see Figure S1b, S1d). The chemistry of these solutions bears a striking resemblance to salts found in Ryugu and Bennu samples (Matsumoto et al., 2024; McCoy et al., 2025). It is intriguing to ponder whether the Bennu parent body could serve as an analog of Enceladus. In addition to hosting Na-rich chloride, carbonate, and phosphate salts, Bennu samples also contain abundant ammonia, methylamine, and acetic acid (Glavin et al., 2025)—similar to materials erupted from Enceladus (Khawaja et al., 2019; Peter et al., 2024). However, this analogy does not mean that their detailed speciations should be the same. As an example, Bennu's soluble phosphates contain both Na and Mg (McCoy et al., 2025), while those at Enceladus have been observed to contain only Na as the counterion (Postberg et al., 2023). Materials from Bennu also attest to desiccation, whereas Enceladus currently has a deep ocean of water. Nevertheless, we suggest that it may be fruitful to further explore their geochemical relationships to better understand pathways of water-rock-organic evolution in the outer solar system.

We find that salts shift and may enter the speciation of DIP on Enceladus (Figure S1a, S1c). For example, $NaHPO_4^-$ can be the dominant form of P depending on the concentration of sodium. We also find a salinity range of 5.1-55 g/kg solution, which can be compared to a minimum of ~20‰ to permit whole-ocean convection (Zeng & Jansen, 2021). As a result, it seems premature to conclude that Enceladus's ocean should be stratified (Ames et al., 2025). By combining our speciation with the model of McCleskey et al. (2012), the electrical conductivity of these ocean samples can be estimated (Castillo-Rogez et al., 2022). The predicted range is 0.83-7.5 S/m. These values can aid efforts to interpret magnetic measurements at Enceladus (Saur et al., 2024).

*3.2. Equilibrium exsolution can explain a high pH*

Our new pH range is significantly higher than the current consensus of ~8-9 (Glein & Waite, 2020; Fifer et al., 2022). Those models employed different assumptions, but they both invoked kinetic control of processes linking the ocean and plume compositions. In contrast, Glein et al. (2015) showed that equilibrium processes occurring during volatile transport can lead to the inference of a high pH fluid. However, their pH values (10.8-13.5) may be too high compared with the range implied by the observed phosphate speciation. Here, we seek to determine whether an updated version of their model can produce more consistent results.

Glein et al. (2015) pointed out that water vapor condensation causes other gases to be enriched in the mixture that emerges from the subsurface of Enceladus. Some water vapor freezes out during transport because of a temperature decrease between the liquid water source of the plume and the local environment where jetting occurs (Goguen et al., 2013). The icy cracks connecting Enceladus's ocean and surface can be thought of as giant distillation columns. We can simplify the problem of accounting for the condensation of water vapor in reconstructing the composition of the gas phase in equilibrium with Enceladean ocean water by assuming that more-volatile gases do not condense during transport (see Section 3.3.1.2.). We focus on $CO_2$ here since its partial pressure can be related to pH through carbonate chemistry. The mass balance of $CO_2$ can be written as

$$\dot{n}_{H_2O,LVE} \times \left( CO_2/H_2O \right)_{LVE} \approx \dot{n}_{H_2O,SVE} \times \left( CO_2/H_2O \right)_{SVE} ,\qquad(6)$$

where $\dot{n}_{H_2O}$ represents a flow rate of water vapor in moles per second. Our notation defines two key regions of equilibrium (Figure 3). Liquid-vapor equilibrium (LVE) corresponds to the region where a liquid water solution is outgassed. Exsolved gases then traverse through cracks in Enceladus's ice shell until a final point of solid-vapor equilibrium (SVE) is reached before the gases erupt at the surface. Equation (6) is a corrected version of Glein et al.'s (2015) mass balance from Fifer et al. (2022). The latter authors pointed out that the flow rate rather than the density of water vapor should serve as the scaling factor of the mixing ratio, as the cross-sectional area and velocity of the flow most likely change between the bottom and top of gas-filled cracks. The molar ratio of gaseous $CO_2/H_2O$ at the bottom is given by

$$\left(CO_2/H_2O\right)_{LVE} \approx \frac{\dot{n}_{H_2O,SVE}}{\dot{n}_{H_2O,LVE}} \times \left(CO_2/H_2O\right)_{SVE} . \tag{7}$$

If it is then recognized that

$$\left(CO_2/H_2O\right)_{LVE} = \frac{p_{CO_2,LVE}}{p_{H_2O,LVE}} , \tag{8}$$

one can derive an equation for the partial pressure of $CO_2$ at the liquid-vapor interface of the crack:

$$\log p_{CO_2,LVE} \approx \log p_{H_2O,LVE} + \log\left(\frac{\dot{n}_{H_2O,SVE}}{\dot{n}_{H_2O,LVE}}\right) + \log\left(CO_2/H_2O\right)_{SVE}. \tag{9}$$

This equation can be evaluated by assuming that the partial pressure of $H_2O$ at the point of LVE is the triple-point pressure of ~0.0061 bar, and the $CO_2/H_2O$ ratio is quenched at this point so that the plume value (~0.005; Peter et al., 2024) can be used. However, we also need to know the fraction of water vapor that condenses between the bottom and top of gas-filled cracks.

**Figure 3.** Artist's impression of how ocean water erupts through cracks on Enceladus. This image highlights processes that affect compositional signatures of erupted materials. Our representation follows from others in the literature (e.g., Spencer et al., 2018; Khawaja et al., 2019), although ours adds details that are clarified in the present paper. By understanding key processes of fractionation, we can properly interpret chemical observations and use them to constrain the chemistry of the input ocean water. Dimensions are not to scale. Credit: A.J. Galaviz (SwRI).

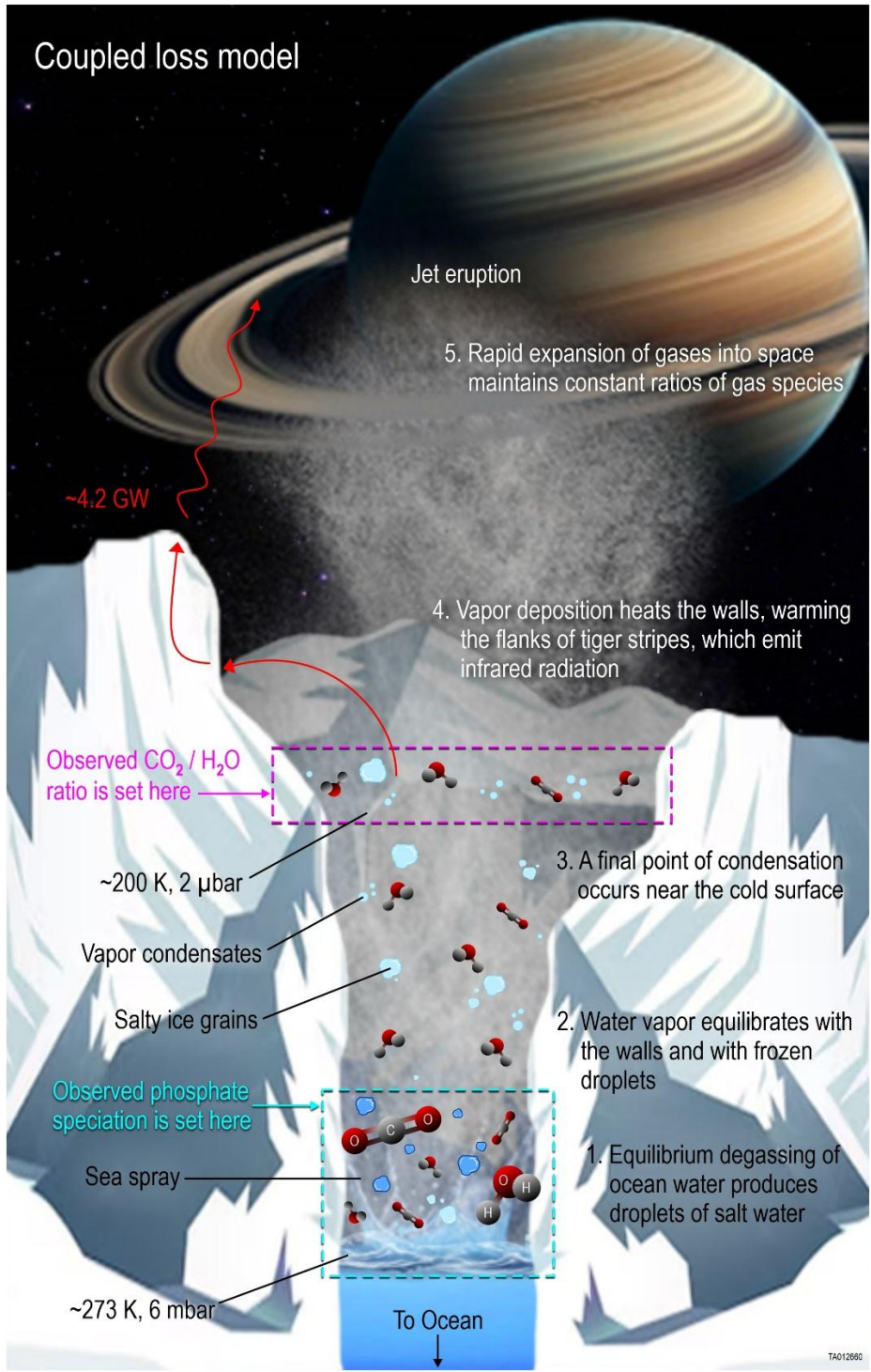

Fifer et al. (2022) address the problem of water vapor condensation by implementing a complex model of heat and mass transfer from Nakajima & Ingersoll (2016). Their model tracks the flow of water vapor in gas-filled cracks. In this work, we opt for a much simpler approach. We will first consider an endmember in which no water vapor condenses to derive an upper limit on $p_{CO_2,LVE}$ (see below). Then, in a later section (see Section 3.3.1.2), we will show that an energy conservation argument can be made to circumvent modeling of fluid dynamics, which depend on unknown details of crack geometry such as depth and width (Nakajima & Ingersoll, 2016). From the perspective of volatile fractionation, such modeling is unnecessary because all we need to know is the fraction of water vapor that condenses, which reflects initial and final states in the crack. We are not questioning the accuracy of Nakajima & Ingersoll's (2016) model compared to ours, and indeed, we expect to find a similar condensed fraction as they did since both models must satisfy the same constraints on the fluxes of water vapor and heat (e.g., Spencer et al., 2018; Hansen et al., 2020).

From Section 3.1, geochemical calculations that reproduce the observed phosphate speciation yield a range of $CO_2$ partial pressures from $10^{-7.27}$ to $10^{-4.23}$ bar. We can set an independent upper limit on $p_{CO_2,LVE}$ based on the $CO_2/H_2O$ ratio in the plume gas. We calculate an upper limit of $10^{-4.5}$ bar using Equation (9) with $\dot{n}_{H_2O,SVE}/\dot{n}_{H_2O,LVE} = 1$ (i.e., no condensation of water vapor). This $p_{CO_2,LVE}$ is consistent with the observed phosphate speciation. Condensation of water vapor will lead to lower inferred values of $p_{CO_2,LVE}$ (see Section 3.3.1.2). Because our model of equilibrium between $CO_2$ dissolved in ocean water and $CO_2$ in the overlying gas phase gives a $p_{CO_2,LVE}$ range that agrees with the $CO_2/H_2O$ ratio measured in Enceladus's plume (Peter et al., 2024), the assumption that this equilibrium is reached appears to be supported. Fifer et al. (2022) assumed instead that gas transfer across the liquid-vapor interface is not at a state of equilibrium but is determined by rates of diffusive transport. This assumption led them to infer that Enceladus's ocean water has a much higher $p_{CO_2}$ than the gas phase adjacent to it (Figure 3). However, such disequilibrium would imply a lower pH of ~8-9 that appears to be inconsistent with the observed phosphate speciation. All forms of DIC contribute to $CO_2$ that is outgassed into the plume in our model. In the model of Fifer et al. (2022), outgassed $CO_2$ comes from dissolved $CO_2$ only. The likelihood of equilibrium being reached during outgassing provides an important constraint on the dynamical timescale of the eruptive process (it should not be too fast), which may have consequences for other aspects of the plume's chemistry, such as how different salts are segregated between ice grains (Postberg et al., 2021).

*3.3. Plume pH versus ocean pH*

Does the high pH recorded by phosphate species reflect the bulk ocean of Enceladus, or does some process modify the pH? One process known to raise pH in natural waters on Earth is $CO_2$ degassing. Spring waters commonly degas $CO_2$ when they emerge from aquifers to Earth's surface, where the partial pressure of $CO_2$ is lower than it is underground (e.g., Choi et al., 1998). The surface of Enceladus is effectively a vacuum, so Enceladus's ocean water should degas $CO_2$ if it is brought close to the surface. In fact, $CO_2$ is one of the major plume gases (Peter et al., 2024). Respeciation after the removal of $CO_2$ (but before freezing) could increase the pH to an unknown extent, because $CO_2$ affects carbonate chemistry differently depending on whether it is dissolved in water or present in the gas phase. We need to understand this process to relate the pH determined from erupted droplets (see Section 3.1) to deeper ocean water.

Figure 4 shows what would happen to the pH if solutions starting with various DIC concentrations and pH values were to undergo $CO_2$ removal. It can be seen that pH does indeed increase. It increases because $CO_2$ is a weak acid in water ("carbonic acid"), and the removal of an acid makes a solution more basic. In terms of the main reactions occurring near the inferred pH range for Enceladus, at first, bicarbonate helps to buffer pH until it is fully "titrated" by $CO_2$ removal (designated by an up arrow, shown below):

$$2HCO_3^- \rightleftharpoons CO_2(\uparrow) + H_2O + CO_3^{-2}; \tag{10}$$

and then, pH increases steeply due to the formation of hydroxide:

$$CO_3^{-2} + H_2O \rightleftharpoons CO_2(\uparrow) + 2OH^-. \tag{11}$$

**Figure 4.** Effects of CO₂ degassing on the pH of a nominal Enceladus salt solution. In each panel, the pH starts at a certain value (7, 8, 9, or 10 at 0 °C and 1 bar), and a family of trajectories can be followed depending on the initial DIC concentration (curves start at 200, 150, 100, or 50 mmolal DIC). The horizontal axis is reversed so that CO₂ is removed from left to right. The blue region demarcates the constrained conditions of final solutions. Triangles signify upper limits on CO₂ loss (and pH) for each curve based on two endmembers of material loss described in Sections 3.3.1 and 3.3.2. A trajectory is allowed if it crosses into the blue region before reaching the triangle of interest. Yellow stars show the amount of CO₂ loss in our preferred scenario of coupled loss of CO₂ and salt solutions, with steam condensation fixed by the observed heat flux.

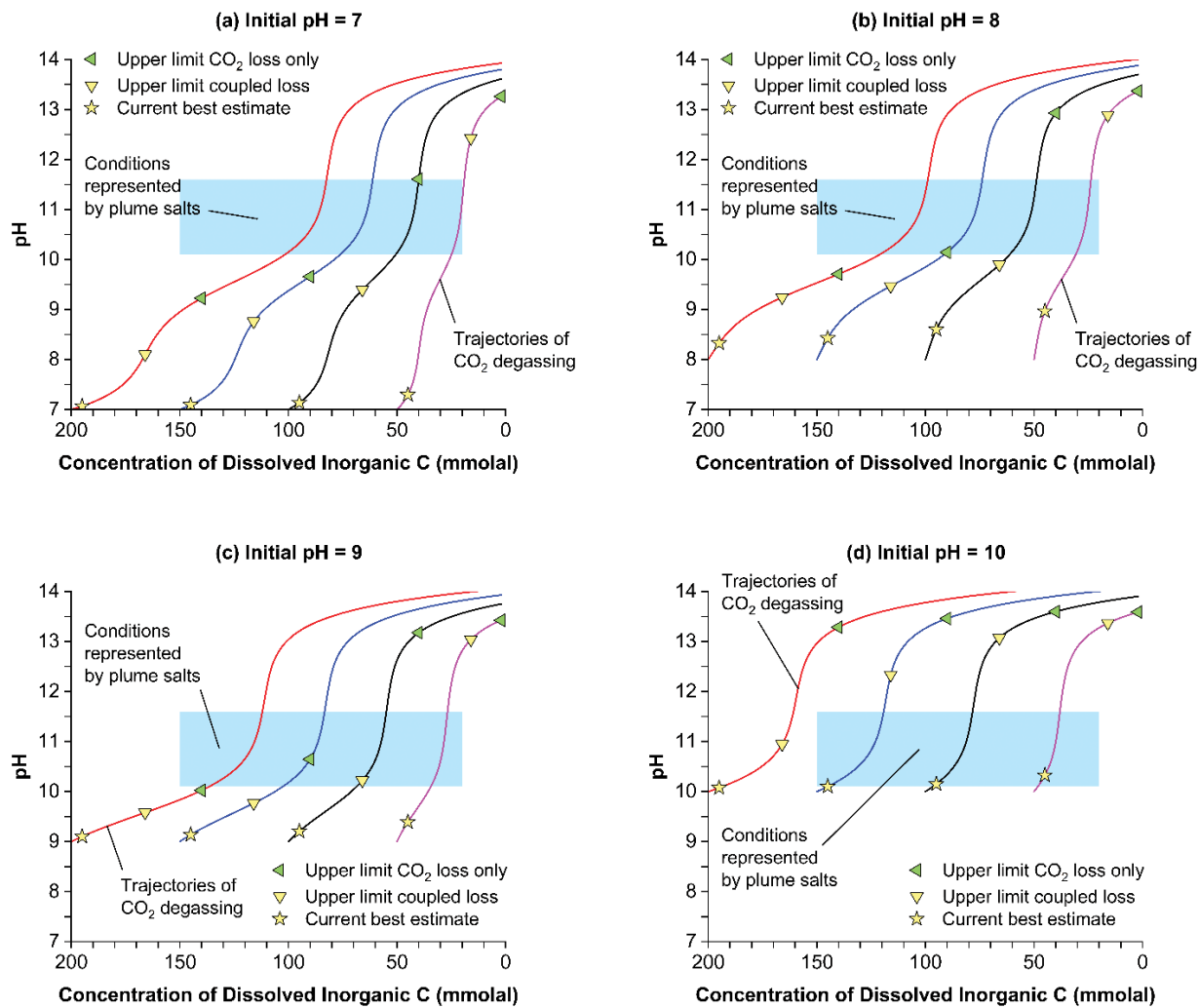

Was the initial pH before degassing significantly lower than the apparent pH? If the amount of degassing is treated as a free parameter, then the pH could have started at values down to 7 (Figure 4a), or possibly lower. Fortunately, we can use the measured abundance of CO₂ as a constraint. We develop two endmember models in an attempt to bound the possibilities. We call these models "perfectly coupled loss" (yellow triangles in Figure 4) and "CO₂ loss only" (green triangles). In the former model, CO₂ is lost together with its host solution. This model (Figure 3) seems to resemble the plume, which contains both gases and salty grains (i.e., frozen host solutions). The erupted material would preserve the bulk composition of the source fluid. However, we cannot dismiss the possibility that gases may be preferentially lost since we do not know *a priori* the dissolved gas content of the source water. While the plume is known to contain salty grains, it could have a larger ratio of $CO_2$-to-salts compared with the ocean below. Thus, it seems wise to also consider an endmember of $CO_2$ loss only. Below, we constrain the loss of $CO_2$ for these scenarios.

### 3.3.1. Perfectly coupled loss of $CO_2$ and salt water

### 3.3.1.1. Most conservative case

If perfectly coupled loss occurs, each kilogram of liquid water going into space would degas $CO_2$ and $H_2O$ into some volume. Because both species enter the same "headspace," their molar quantities ($n$) and partial pressures follow a simple relationship:

$$\frac{n_{CO_2,LVE}}{n_{H_2O,LVE}} = \frac{p_{CO_2,LVE}}{p_{H_2O,LVE}} \,. \tag{12}$$

As described in Section 3.2, $p_{H_2O,LVE} \approx 0.0061$ bar and $p_{CO_2,LVE} \leq 10^{-4.5}$ bar. The amount of water vapor that will be produced can be determined by setting an energy balance between the latent heats of fusion ($\Delta H_{fus}$) and vaporization ($\Delta H_{vap}$) of water at triple-point conditions. The fraction of water vaporized ($f_{vap}$) is therefore given by

$$f_{vap} = \frac{\Delta H_{fus}}{\Delta H_{fus} + \Delta H_{vap}} = \frac{6.01 \text{ kJ/mol}}{6.01 + 45.06} = 0.118 \,. \tag{13}$$

Each kilogram of liquid water contains ~55.5 moles, so ~6.55 moles of water vapor will be produced. Using the above values, we can calculate the amount of $CO_2$ via Equation (12). We find that at most ~34 mmol $CO_2$/kg liquid water could be degassed.

### 3.3.1.2. Current best estimate

The preceding upper limit corresponds to a case with no water vapor condensation in Enceladus's tiger stripes. However, there are grains of almost pure ice being erupted (Postberg et al., 2009) and substantial thermal emission from the tiger stripes (Spencer et al., 2006). The deposition of water from gas to solid can explain both observations (Schmidt et al., 2008; Nakajima & Ingersoll, 2016). The rate of deposition ($D_{H_2O}$) can be calculated via another energy balance that assumes that the radiant flux emitted by the tiger stripes ($\Phi_{tiger}$; Spencer et al., 2018) comes from the deposition of water vapor (i.e., the reverse of sublimation). We find that

$$D_{H_2O} \approx \frac{\Phi_{tiger}}{\Delta H_{sub}} \approx \frac{4.2 \times 10^9 \text{ W}}{2.83 \times 10^6 \text{ J/kg}} \approx 1500 \text{ kg/s} \,. \tag{14}$$

It is then simple to estimate the amount of water vapor remaining in the plume gas compared with the amount originally vaporized from liquid water. The ratio can be expressed as

$$\frac{\dot{n}_{H_2O,SVE}}{\dot{n}_{H_2O,LVE}} = \frac{L_{H_2O}}{L_{H_2O} + D_{H_2O}} \approx \frac{300}{300 + 1500} \approx 0.17 \,, \tag{15}$$

where the loss rate of water vapor ($L_{H_2O}$ in kg/s) is from Hansen et al. (2020). Inserting the above result into Equation (9) yields $p_{CO_2,LVE} \approx 10^{-5.3}$ bar, which corresponds to degassing of ~5 mmol $CO_2$/kg liquid water. This is our preferred scenario. If we speciate the two endmembers of salinity shown in Table S1 with this value of $p_{CO_2,LVE}$, we obtain a pH of 10.6-10.7—remarkably consistent with the range (10.1-11.6) independently estimated from the phosphate speciation (Figure 2).

Note that Fifer et al. (2022) estimated that ~60-70% of water vapor should condense during transport, which is similar (and essentially identical on a logarithmic scale; see Equation 9) to our value of ~83% (see Equation 15). Thus, the large difference between pH values boils down to how the two models treat the liquid-vapor interface. We assume that gases equilibrate between the two phases (or at least get close to equilibrium), whereas Fifer et al. (2022) assumed that transport from the liquid to the gas phase is kinetically controlled.

$CO_2$ condensation (Combe et al., 2019) would lead to a higher model-derived pH compared with the actual pH (Glein et al., 2015). However, this process may not significantly affect our derivation of pH because only ~83% of water vapor is expected to freeze out, and $CO_2$ is far less condensable than $H_2O$ is. For context, $CO_2$ ice has a vapor pressure that is larger than that of $H_2O$ ice by a factor of $10^6$ at 200 K (Huber et al., 2022). This temperature is comparable to that (177-217 K) determined by Goguen et al. (2013) after Cassini's infrared mapper peered into one of Enceladus's tiger stripes (Baghdad Sulcus). It might serve as a reasonable estimate for the final point of solid-vapor equilibrium prior to plume formation (Figure 3). We suspect that only a small fraction (<1%) of $CO_2$ may condense inside Enceladus's tiger stripes because they are too warm.

### 3.3.2. $CO_2$ loss with retention of salt water inside Enceladus

Alternatively, we can consider Enceladus's ocean to be a finite reservoir that simply loses $CO_2$, causing the remaining DIC concentration to decrease through time. The change in DIC molality due to outgassing can be related to the loss rate of $CO_2$ from the plume ($L_{CO_2} \approx 2.6 \times 10^9$ mol/yr; Hansen et al., 2020; Peter et al., 2024), as shown below:

$$\Delta m_{DIC} \approx -\frac{L_{CO_2} \tau_{out}}{M_{oc}}, \qquad (16)$$

where $M_{oc}$ stands for the mass of the ocean (~1.3×10$^{19}$ kg; Park et al., 2024), and $\tau_{out}$ designates the outgassing timescale. Hörst et al. (2008) proposed that CO in Titan's atmosphere is sourced by O$^+$ ultimately derived from water in Enceladus's plume. They calculated that it would take ~300 Myr for CO to build up to the level now observed. We suggest that this process could date the duration of plume activity (it may also provide a constraint on the age of Enceladus; see Nimmo et al., 2023), giving us $\Delta m_{DIC} \approx -60$ mmol $CO_2$/kg liquid water from Equation (16). However, this value is probably an upper limit on $CO_2$ loss for two reasons: (1) Carbonate minerals on the seafloor of Enceladus should dissolve within this timeframe (Hao et al., 2022), helping to restore DIC; and (2) If some of Titan's CO is primordial or was delivered by comets, then less CO production would be implied.

### 3.3.3. Synthesis: Insignificant evolution of pH

The above constraints on the amount of $CO_2$ degassing can be used to restrict the parameter space of pH-DIC conditions of pristine ocean water. The fewest restrictions come from our model of $CO_2$ loss only. In this case, the initial pH could be 7 (or lower) but only if the initial DIC concentration is less than ~120 mmolal (Figure 4a). The initial pH can be 10 (or higher) provided that a trajectory is chosen that ends inside the blue box (Figure 4d). However, "$CO_2$ loss only" may not be realistic (Figure 3). Ocean water with a pH of 10 is susceptible to a big jump in pH with the loss of $CO_2$ because most (~60%) of its DIC is in the forms of $NaCO_3^-$ and $CO_3^{-2}$. These carbonate species can easily generate a significant amount of $OH^-$. In contrast, starting solutions with lower pH values and high DIC concentrations, in particular, have more base-buffering capacity. Assuming that "coupled loss" is more realistic, then an initial pH of 7 would be inconsistent with the inferred pH range unless the initial DIC concentration is relatively low (<70 mmolal; Figure 4a). Yet, such a low DIC concentration may not be sufficient to drive the release of abundant phosphate from minerals (Postberg et al., 2023). While we cannot rule out a circumneutral pH for Enceladus's ocean, we consider it unlikely. Our preferred model of outgassing (indicated by stars in Figure 4) predicts small changes in DIC concentration and pH. A minimal extent (<0.2 units) of pH evolution during degassing means that the pH of plume salt solutions should be representative of pristine ocean water (see Section 4.3).

## 4. Implications

### 4.1. Water-rock interactions create alkaline ocean water

Our new understanding of the pH of Enceladus's ocean and how volatile signatures are transferred between the ocean and plume have important implications. High pH ocean water is evidence of a strong degree of interaction between chemically basic rocks and the ocean of Enceladus (Glein et al., 2015). Aqueous alteration (e.g., serpentinization) of Mg- and Fe(II)-rich silicates, such as those found in carbonaceous asteroids (Nakamura et al., 2023; Lauretta et al., 2024), is one possibility. Another possibility is the alteration of peralkaline rocks with molar Na/Al > 1. Refractory cometary grains appear to be peralkaline (Bardyn et al., 2017). The alteration of such material would release excess NaOH into solution, which after reacting with $CO_2$, may be responsible for a high alkalinity of sodium carbonates (Postberg et al., 2009; 2021).

If Enceladus's core contains rocks similar to dust observed from comet 67P, then this bulk composition may allow much higher concentrations of DIP to exist than those predicted for chemical equilibrium between liquid water and chondritic rocks (Randolph-Flagg et al., 2023). A key distinction is the Na/Al ratio, which is below unity in carbonaceous chondrites (Lodders, 2021). In saponite (e.g., $Na_{0.33}Mg_3Si_{3.67}Al_{0.33}O_{10}(OH)_2$), an abundant secondary mineral in aqueously altered chondrites (Lee et al., 2025), substitution of Si by Al in tetrahedral sites enables a corresponding amount of Na to be incorporated into the interlayer space. This incorporation makes it more difficult to release Na into aqueous solution. When the rock is peralkaline, however, Na cannot be completely charge-balanced by Al in stable aluminosilicates. Thus, Na is more prone to leaching and can go on to react with accreted $CO_2$, producing a solution with high pH and high carbonate alkalinity. The latter quantity increases the solubilities of phosphate minerals (Hao et al., 2022). We note that these are our expectations, and that this type of alteration has not been modeled yet for Enceladus's ocean.

### 4.2. How strong is the evidence for hydrothermal silica?

A second implication of this work is that a high ocean pH adds to a body of evidence that may question the interpretation of silica ($SiO_2$) nanoparticles as indicators of the cooling of hydrothermal fluids inside Enceladus (Hsu et al., 2015). It was argued that forming and maintaining these particles requires a pH between 8.5 and 10.5. At higher pH, colloidal silica is too soluble. Our estimated pH range (10.1-11.6) is marginally consistent with the proposed solubility limit. The hypothesis of hydrothermal silica also has other problems. First, it is challenging to find conditions that permit the coexistence of high $SiO_2$ and $H_2$ abundances. Silica inhibits $H_2$ production by stabilizing ferrous iron in silicate minerals, rather than supporting the oxidation of iron by water (e.g., Sleep et al., 2004). Chemical heterogeneity of Enceladus's core may need to be invoked to reconcile current interpretations for $SiO_2$ and $H_2$ (Glein et al., 2018). Finally, modeling of ocean circulation suggests that transport times from the bottom to the top of Enceladus's ocean are probably too long to allow silica nanoparticles to survive this passage (Zeng & Jansen, 2021; Kang et al., 2022; Ames et al., 2025; see Schoenfeld et al., 2023; Bouffard et al., 2025). It seems hard to understand how such particles would not grow/dissolve before being embedded in plume ice grains. These findings highlight the importance of exploring alternative explanations of the silica nanoparticle data in the future.

*4.3. Reconstruction of ocean chemistry from plume composition*

Since the phosphate-derived pH provides a validation of our outgassing model, we now have an improved tool (see Waite et al., 2017; Fifer et al., 2022; Mitchell et al., 2024) for converting between plume gas mixing ratios and dissolved concentrations in Enceladus's ocean. As an example, here we apply our model to $H_2$, $CH_4$, and $NH_3$. To reconstruct the composition of the ocean, we conduct three sets of calculations. In the first set, we determine concentrations of gases that are still dissolved after degassing. While we could just use Henry's law for $H_2$ and $CH_4$, a full speciation model must be run because $NH_3$ will be in equilibrium with $NH_4^+$. We adopt nominal values for the concentrations of Cl (0.2 molal), DIC (0.1 molal), and DIP (0.004 molal), as well as previously used ratios of K/Na and Br/Cl. The pH (~10.7) is determined via charge balance based on the partial pressure of $CO_2$. We impose fixed partial pressures of gases from Equation (9) with $\dot{n}_{H_2O,SVE}/\dot{n}_{H_2O,LVE} \approx 0.17$ (see Equation 15) and observed mixing ratios (Waite et al., 2017; Peter et al., 2024). We speciate this system using React in GWB (with decoupled redox species) and find the following dissolved concentrations: $H_2 = 1\times10^{-8}$ molal, $CH_4 = 3\times10^{-9}$ molal, and $\Sigma NH_3 = 0.006$ molal ($\Sigma NH_3$ stands for $NH_3(aq) + NH_4^+$).

Next, we calculate the degassed amounts of $H_2$, $CH_4$, and $NH_3$. A suitable equation can be derived by multiplying Equation (7) by the number of moles of water vapor also generated during evaporative freezing, and generalizing the result:

$$\frac{n_{Gas,LVE}}{kg\ liquid} \approx \frac{n_{H_2O,LVE}}{kg\ liquid} \times \frac{\dot{n}_{H_2O,SVE}}{\dot{n}_{H_2O,LVE}} \times \left(Gas/H_2O\right)_{SVE}, \tag{17}$$

where "Gas" refers to any non-condensable gas, and Gas/$H_2O$ designates the molar ratio of this species with respect to water vapor. The left side of Equation (17) signifies the molality of gas removed. In Section 3.3.1.1, we showed that $n_{H_2O,LVE}$ could be ~6.55 moles per kilogram of liquid water. We can also adopt a value of ~0.17 for $\dot{n}_{H_2O,SVE}/\dot{n}_{H_2O,LVE}$ (see Equation 15). Thus,

$$\frac{mol\ Gas}{kg\ liquid} \approx 1.1\times\left(Gas/H_2O\right)_{SVE}. \tag{18}$$

Based on the plume gas composition (Waite et al., 2017; Peter et al., 2024), the degassed amounts are calculated to be: $H_2 = 0.01$ molal, $CH_4 = 0.001$ molal, and $NH_3 = 0.02$ molal.

The final step is to add the outgassed inventory back into the previously speciated solution and respeciate the new solution with the full volatile inventory. Such "regassing" can be simulated in React using the pickup command. We also add in the outgassed amounts of $CO_2$ (~0.005 mol/kg liquid water; see Section 3.3.1.2) and $H_2O$ (~6.55 mol/kg liquid water; see Section 3.3.1.1). Table A1 displays the calculated ocean speciation. This output represents a nominal estimate of the inorganic chemistry of Enceladus's ocean before ocean water experiences compositional modification due to the eruptive process that forms the plume. It is tempting to call this "the ocean composition," and it may be, although a detailed analysis of the uncertainty space propagating from the input data has yet to be performed. The main point of this exercise is to demonstrate how the chemistry of the ocean can be inferred using a framework that is consistent with constraints from observations and the thermodynamics of outgassing on Enceladus. Interestingly, regassing hardly decreases the pH (<0.1 units). $CO_2$ degassing has the tendency to increase pH but is partly counterbalanced by $NH_3$ degassing. They have opposing acid-base effects. This is even better news than our finding from Section 3.3.3 because it makes us more confident that future measurements of the pH of plume ice (after collection and melting) will be representative of the underlying ocean.

Lastly, this nominal ocean composition (Table A1) can serve as a baseline for future studies, and it may motivate us to revisit some previous assumptions about the geochemistry and habitability of Enceladus. For example, could its ocean contain too much free $NH_3$ for life (McKay et al., 2008)? Are key transition metals that are needed to produce certain biological catalysts (e.g., hydrogenases) sufficiently abundant at this high pH (Xu et al., 2025)? Could $NH_3$ also contribute to the high pH (e.g., Leitner & Lunine, 2019)? Although our estimated pH is about two units higher than that of Fifer et al. (2022), our estimates of dissolved gas concentrations in Enceladus's ocean also support the existence of a gassy ocean, as originally suggested by Kargel (2006). However, we find that $H_2$ could be the dominant low-solubility gas, and free $CO_2$ may not be abundant as the pH is high. A high $H_2$ concentration drives the $CO_2$-$CH_4$ redox couple out of chemical equilibrium, and we calculate that hydrogenotrophic methanogenesis could supply ~140 kJ/mol $CH_4$ of free energy below the zone of degassing in the ocean. Enceladus's ocean remains energetically habitable (Waite et al., 2017; Fifer et al., 2022). The high abundance of $H_2$ may also be geophysically relevant. Mitchell et al. (2024) emphasized that $H_2$ exsolution could drive explosive cryovolcanism on Enceladus. Our geochemical model suggests that $H_2$ would exsolve at ~10 bars of total pressure, or at depths shallower than ~9.6 km in the ice shell of Enceladus, thus supporting a role for $H_2$ in Enceladus's eruption dynamics. Table A1 may incite numerous additional questions, which can help to catalyze progress in achieving a deeper understanding of Enceladus. The future looks bright. Moreover, the general capability of our model for dissolved gases may prove useful in interpreting current/future measurements of other volatiles, including certain organic compounds (e.g., MacKenzie et al., 2021).

## Acknowledgments

This work was supported by NASA's Habitable Worlds program. We thank Frank Postberg for making his group's updated plume salt numbers available to the community by presenting them at the 2021 AGU Fall Meeting. We are also grateful to two reviewers who made a number of excellent suggestions for improvements.

**Appendix A. Geochemical properties of Enceladean ocean water prior to degassing**

Table A1 shows results from a series of thermodynamic calculations that are described in Section 4.3. In the laboratory, this speciation can be produced by preparing an aqueous solution containing 0.158 molal NaCl, 0.0180 molal KCl, 0.0722 molal $Na_2CO_3$, 0.0210 molal $NaHCO_3$, 0.00353 molal $Na_2HPO_4$, 0.000404 molal NaBr, and 0.0255 molal $NH_4OH$ under a 10.6 bar atmosphere containing 96 mol % $H_2$ and 4 mol % $CH_4$ at 0 °C. One could test whether methanogens can grow in Enceladus's ocean.

**Table A1.** Nominal geochemical properties of Enceladus's ocean at 0 °C and 1 bar. Brackets indicate the concentration of the enclosed species. Molal concentrations are given down to a value of $1\times10^{-9}$. Organic species (Khawaja et al., 2019) are not included.

| Property | Value |
| --- | --- |
| pH | 10.6 |
| Eh (V) [a] | -0.621 |
| $\log f_{O_2}$ (bar) [b] | -94.3 |
| $\log f_{O_2}$ (vs. FMQ) [c] | -5.9 |
| $p_{H_2}$ (bar) [d] | 10.2 |
| $p_{CH_4}$ (bar) [d] | 0.4 |
| Ionic strength (molal) | 0.333 |
| Activity of $H_2O$ | 0.994 |
| $A_{CO_2-CH_4}$ (kJ/mol $CH_4$) [e] | 140 |
| [$Na^+$] | 0.274 |
| [$Cl^-$] | 0.173 |
| [$NaCO_3^-$] | 0.0482 |
| [$CO_3^{-2}$] | 0.0311 |
| [$NH_3(aq)$] [f] | 0.0181 |
| [$K^+$] | 0.0179 |
| [$HCO_3^-$] | 0.0106 |
| [$H_2(aq)$] [g] | 0.0100 |
| [$NH_4^+$] | 0.00744 |
| [$NaCl(aq)$] | 0.00335 |
| [$NaHCO_3(aq)$] | 0.00328 |
| [$NaHPO_4^-$] | 0.00156 |
| [$HPO_4^{-2}$] | 0.00147 |
| [$CH_4(aq)$] [f] | 0.00100 |
| [$Br^-$] | 0.000402 |
| [$Na_2HPO_4(aq)$] | 0.000177 |
| [$PO_4^{-3}$] | 0.000101 |
| [$NaPO_4^{-2}$] | 7.55e-05 |
| [$Na_2PO_4^-$] | 7.36e-05 |
| [$OH^-$] | 7.29e-05 |
| [$KHPO_4^-$] | 5.88e-05 |
| [$NaOH(aq)$] | 6.27e-06 |
| [$KPO_4^{-2}$] | 4.11e-06 |
| [$KCl(aq)$] | 2.04e-06 |
| [$NaBr(aq)$] | 1.83e-06 |
| [$K_2HPO_4(aq)$] | 1.52e-06 |
| [CO$_2$(aq)] [f] | 6.49e-07 |
| [H$_2$PO$_4^-$] | 2.33e-07 |
| [KOH(aq)] | 1.68e-07 |
| [K$_2$PO$_4^-$] | 7.57e-08 |
| [NaH$_2$PO$_4$(aq)] | 6.39e-08 |
| [KBr(aq)] | 3.70e-08 |
| [KH$_2$PO$_4$(aq)] | 5.91e-09 |
| [P$_2$O$_7^{-4}$] | 2.29e-09 |
| [H$_2$CO$_3$(aq)] | 1.10e-09 |

[a] Reduction potential relative to standard hydrogen electrode.

[b] Oxygen fugacity.

[c] Fayalite-magnetite-quartz buffer based on thermodynamic data from Holland & Powell (2011).

[d] Equilibrium partial pressure.

[e] Chemical affinity for CO$_2$(aq) + 4H$_2$(aq) → CH$_4$(aq) + 2H$_2$O(liq), which quantifies how far this reaction is from chemical equilibrium.

[f] Dissolved concentration based on plume gas data from Peter et al. (2024).

[g] Dissolved concentration based on plume gas data from Waite et al. (2017).